\RequirePackage{fix-cm}
\documentclass[twocolumn]{svjour3}          
\smartqed  
\usepackage{graphicx}
\usepackage{epstopdf}

\begin{document}

\title{Effect of Ti underlayer thickness on the magnetic anisotropy of TbFe thin
films}

\author { {Ajit Kumar Sahoo}       \and J. Arout Chelvane \and
        J. Mohanty
}


\institute{Ajit Kumar Sahoo \at
              Nanomagnetism and Microscopy Laboratory, Department of Physics\\ Indian Institute of Technology Hyderabad, Kandi, Sangareddy, Telangana, 502285, India \\
              \email{ph17resch11011@iith.ac.in}           
           \and
           J. Arout Chelvane \at
           Defence Metallurgical Research Laboratory, Hyderabad, 500058, India\\
           \email{aroutchelvane@gmail.com}
           \and
           J. Mohanty \at
              Nanomagnetism and Microscopy Laboratory, Department of Physics\\ Indian Institute of Technology Hyderabad, Kandi, Sangareddy, Telangana, 502285, India \\
              \email{jmohanty@phy.iith.ac.in}
}

\date{Received: \today / Accepted: date}
\maketitle

\begin{abstract}

In this study, we address the impact of Ti underlayer thickness
(UL: 0-40~nm) on the structural, magnetic, and microscopic
properties of TbFe thin films. The structural analysis confirmed
the intermixing at interfaces of the Ti and TbFe layer with the
increment of UL thicknesses. Out-of-plane (OOP) coercivity
($H_c$), and saturation field ($H_s$) gradually increased with an
increase in UL thickness regardless of interface mixing. For UL =
10~nm, the domain contrast and OOP stray field strength were
enhanced, which may be due to the extent of $d$-$d$ hybridization
dominated over the influence of interfacial roughness. While for
UL = 20, and 40 nm, the extent of interfacial roughness dominated
the hybridization effects and as a result, stray fields
deteriorated. By placing UL of 20 nm, $H_c$ increased by nearly 6
times more than the bare TbFe system. So, we observe a state with
high OOP $H_c$ combined with nearly zero OOP stray fields that are
found to co-exist in the sample. The magnetization reversal
studies on a large area reveal domain nucleation followed by
domain-wall motion in all the films. The idea of tuning magnetic
properties by varying thicknesses of Ti UL may useful in
spintronics applications.


\end{abstract}

\section{Introduction}
\label{intro}

The improvement of magnetic properties of thin films associated
with perpendicular magnetic anisotropy (PMA) system has been
emphasized, for many decades, due to the possible industrial
applications like high-density perpendicular recording media,
magnetic tunnel junctions, sensors, etc. The PMA systems like
CoPt, FePt alloys are crystalline, therefore tuning of PMA can be
studied by understanding the magneto-crystalline anisotropy in the
system~\cite{CoPt,FePt,FePt_cap}. However, the theory behind PMA
in amorphous systems like rare-earth transition-metal (RE-TM),
CoFeB, CoCrPt, is still elusive till date
~\cite{GdFe_amor,TbFe_amor,CoFeB_amor,CoCrPt_amor}. The PMA in
these systems can be developed by simply varying thickness,
strain-modulation, and interface inducing roughness in the film,
etc. Several approaches have been considered for further
improvement in PMA property, by deposition of a suitable
underlayer (UL) or seed-layer, incremental in the repetition of
multilayers (ML), inducing more interfacial roughness, etc.

Literature suggests various UL like Ti, V, Cr, Ni, Cu, Ta, Pd,
etc. have been tested on various films like FeCo, FePt, CoCrPt,
CoCr, FePt ML, etc. For example, Ti UL was used to improve the
degree of order of FePt films~\cite{FePt}, to improve the PMA
property of CoCrPt~\cite{CoCrPt_amor} and CoCr \cite{CoCr} films.
Interestingly it was observed that Ti and Ta UL made considerable
improvement in magnetic properties of FeCo films compared to other
UL~\cite{FeCo}. This may be due to the development of grain size
and the internal stress of the film. Another important thing is,
Ti UL also minimizes the contact resistance of
single-wall-nano-tube and a metal electrode which can be
beneficial for various electrical applications~\cite{SWDW}.

RE-TM films exhibit an impressive PMA property suitable for
possible spintronics applications. GdFe~\cite{GdFe_amor}, TbFe
~\cite{TbFe_amor}, GdFeCo~\cite{GdFeCo} many more systems were
investigated on the light of their thickness, composition, and
various UL to enhance PMA property. This tuning of properties
depends strictly on the thickness of the film and interface
properties. When the film thickness ~$<$ $5$~nm, interface-driven
properties or effects become dominant. However, for $\geq$ $5$~nm
volume thickness shows an effective bulk-like property, whereas
interface effects become negligible. These bulk properties are
mostly observed in the amorphous system irrespective of the sample
preparation technique.

Here, we have chosen an amorphous ferrimagnetic TbFe system with
various thicknesses of Ti UL to observe a full extent of variance
in structural, magnetic, and microscopic properties. First, we
observed TbFe of 40~nm thickness gives a clear PMA sign with nice
stripe magnetic domains. Then, we varied the Ti layer of various
thicknesses (10, 20, and 40~nm) to observe the effect of interface
roughness, pinning sites, and hybridization effects of the films
towards structural, magnetic, and microscopic phenomena. Here, we
observed higher OOP $H_c$ values along with nearly null OOP stray
fields in UL of 20~nm and 40~nm sample. This observation is quite
new in these kinds of bulk PMA system, as of our knowledge.

\section{Experimental detail}

Si$<1 0 0>$ / Ti($t_{UL}$) / TbFe(40~nm) / Cr(3~nm); with $t_{UL}$
= 0, 10, 20, and 40~nm thin films were prepared by electron beam
evaporation technique. Cr (3~nm) was used as a capping layer to
protect the film from oxidation, while Ti is used as UL. Films
were deposited with a rate of deposition $\approx 3$ \AA/sec.
Tb-Fe films were deposited from its composite alloy, after
deposition the composition of the film was estimated by energy
dispersive spectroscopy (EDS) analysis. For $t_{UL}$ = 0~nm can be
noted as a bare TbFe film of 40~nm, which means TbFe film does not
have any UL influence. The background pressure was maintained at
around $2\times 10^{-6}$ $Torr$ during deposition. To understand
the effect of various thicknesses of UL on structural formations
grazing incident X-ray diffraction (GI-XRD), and the X-ray
reflectivity (XRR) technique were used. Atomic force microscopy
(AFM) was used for identifying the topographic information.
Magnetic force microscopy (MFM) was used to probe the stray fields
emanating from the film surface. For hysteresis and domain
reversal studies, we have used polar magneto-optical Kerr
microscopy (PMOKE) (spot size $\approx$ 3 $\mu$m).

\section{Results and discussions}

\begin{figure}[ht]
\centering
\includegraphics[scale=0.5]{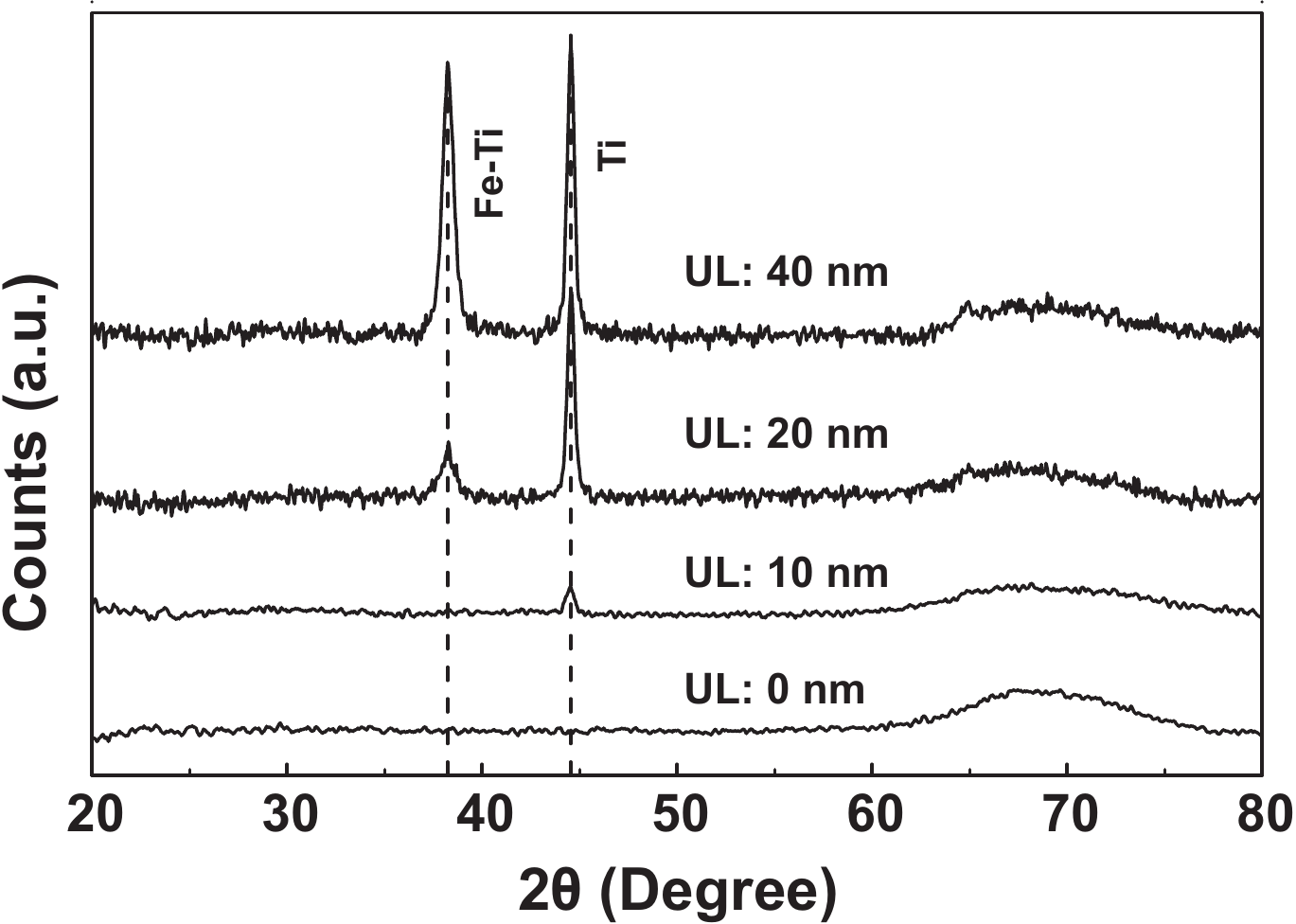}
\caption{GIXRD patterns of Tb-Fe film with various thicknesses of
Ti UL (0, 10, 20, and 40~nm).}
\label{fig:1}
\end{figure}

\begin{figure}[ht]
\centering
\includegraphics[scale=0.35]{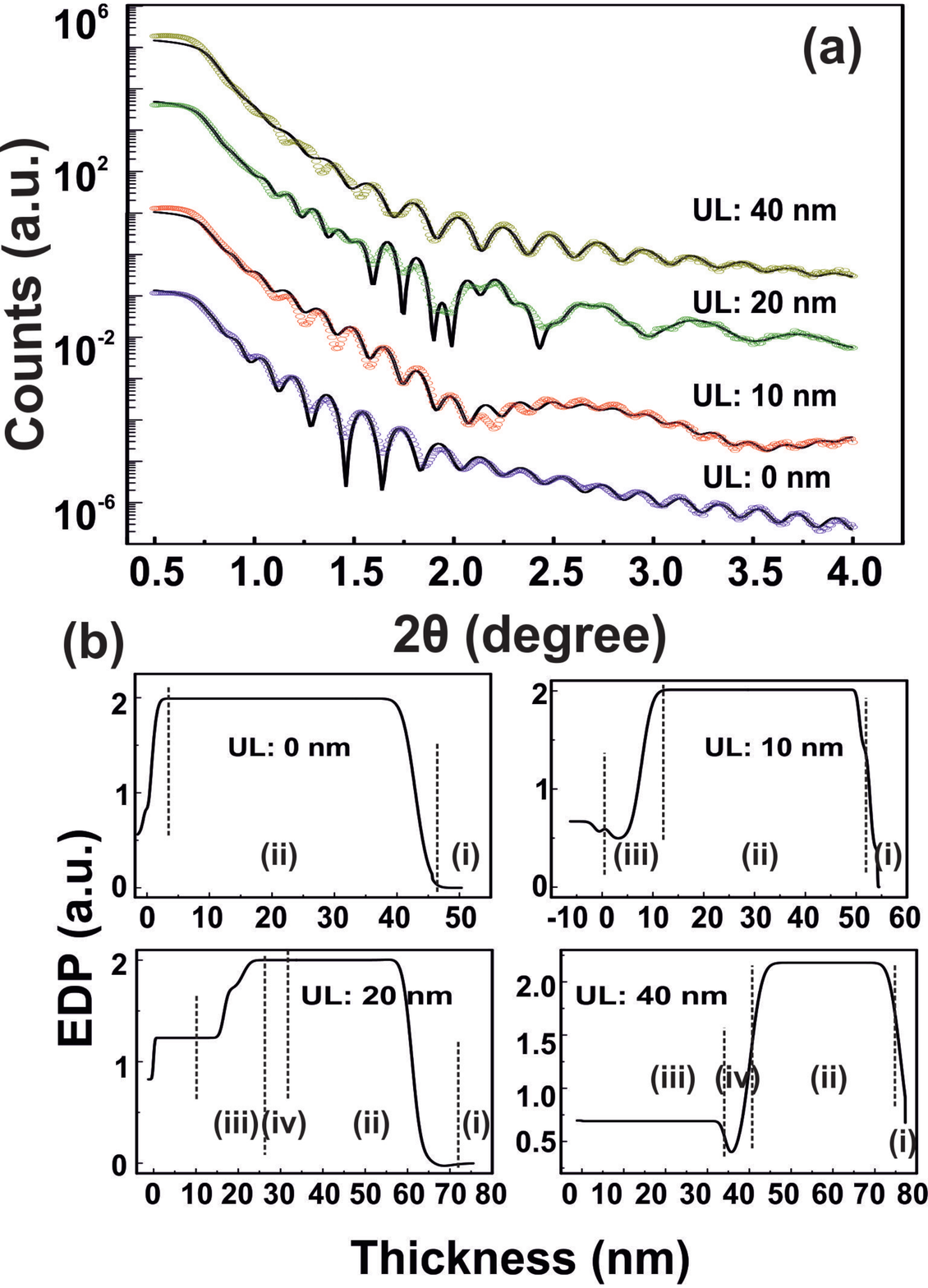}
\caption{(color online) (a) X-ray reflectivity spectra as a
function of 2$\theta$ of various films, UL: 0, 10, 20, and 40 nm.
(b) Corresponding electron density profile (EDP) of various films.
The region$-$(i), (ii), (iii), and (iv) represent Cr, TbFe, Ti UL,
TiFe inter-diffusion layers.}
\label{fig:2}
\end{figure}

From EDS analysis (not presented here) composition of the film
turns out to be $\approx$ $Tb_{30}Fe_{70}$ for all the samples.
Figure~\ref{fig:1} represents GIXRD patterns of the Tb-Fe film
with various UL thicknesses. Bare 40~nm  thick TbFe film of shows
an amorphous signature, which is consistent with amorphous nature
of RE-TM
~\cite{RE_TM,GdFe_amorphous,TbFeCo_amorphous,DyTbFeCo_amorphous,TbFe_oPMA_amorphous}.
However, it is observed in akin GdFe films with the Gd composition
less than 15 at. \% are not amorphous, while more than 20 at. \%
shows an amorphous behavior~\cite{GdFe_per_amor}. Comparing with
the literature database, in this case, Tb composition was
estimated as $\approx$ 30 at. \% for all the TbFe films, therefore
it is expected to exhibit an amorphous nature in the as-deposited
state. For UL of 10 nm, the film shows the hcp phase of Ti, which
is at $2\theta$ $\approx$ 44$^{\circ}$. Along with this, the Fe-Ti
phase is observed in the case of UL = 20, and 40 nm. This means
that the possibility of diffusion of Fe and Ti may occur at the
interface of Ti-UL and TbFe film. The effect of this diffusion
during the growth gets enhanced when UL thicknesses are increased
which is also evident from the peak intensity. To further probe
the individual layer thickness and rms roughness we implement a
non-destructive XRR technique. Figure~\ref{fig:2} (a) and
~\ref{fig:2} (b) represents XRR spectra and corresponding electron
density profile (EDP) of various films respectively. Despite of
being amorphous in the case of UL: 0 nm (observed by GIXRD), we
observe Kiessig fringes in the corresponding reflectivity spectra
because of the difference in electron density of TbFe and Cr
layers. The extracted data of individual layer thickness (t), root
mean square (rms) roughness ($\sigma$), and density ($\rho$) of
the layer are summarized in Table~\ref{tab1}. The composition of
$Tb_{0.3}Fe_{0.7}$ well agreed with the experimental spectra,
which is also evident from EDS analysis earlier. Expected
thickness of TbFe of $40$ nm and Cr of $3$ nm (UL: 0 nm) is found
to be as $42.5$ nm and $2.7$ nm respectively from XRR analysis.
Slight increment in inter-diffusion of the TbFe and Ti layer
(formed as $Ti_{0.3}Fe_{0.7}$) is observed for increase in UL
thickness. Similar evidence of UL (Cu) diffusion into the RE-TM
magnetic layer (Sm$Co_5$) is observed using the elemental map and
energy dispersive x-ray spectroscopy line analysis
~\cite{underlayer_Cu}. The increment of UL thickness has a
significant impact on the microstructural, magnetic, and
microscopic properties of the sample.

\begin{table}[ht]
\centering
\caption{The extracted parameters thickness ($t$ in
$nm$), rms interfacial roughness ($\sigma$ in $nm$), and density
($\rho$ in $g/cm^{3}$) of various layers in sample with UL
thickness: 0, 10, 20, and 40 $nm$ are summarized here.}

    \begin{tabular}{|p{0.9cm}|p{1.3cm}|p{0.5cm}|p{1.3cm}|p{1.3cm}|p{0.5cm}|}
    \hline
    Sample & Parameter & Ti    & $Ti_{0.3}Fe_{0.7}$ & $Tb_{0.3}Fe_{0.7}$ & Cr \\
    \hline
          & t ($\pm 0.5$) &     -  &    -   & 42.5  & 2.7 \\

    0 nm & $\sigma$ ($\pm 0.1$) &    -   &    -   & 1.2   & 1.1 \\

          & $\rho$ ($\pm 0.01$) &   -    &   -    & 8.42   & 6.04 \\
    \hline
          & t ($\pm 0.5$) & 7.9   &  -     & 42.9  & 2.5 \\

    10 nm & $\sigma$ ($\pm 0.1$) & 1.1   &   -    & 0.6  & 0.4 \\

          & $\rho$ ($\pm 0.01$) & 4.75  &   -    & 8.50   & 5.18 \\
    \hline
          & t ($\pm 0.5$) & 16.7  & 4.5   & 39.9  & 3.4 \\

    20 nm & $\sigma$ ($\pm 0.1$) & 1.5   & 1.4   & 1.7   & 1.3 \\

          & $\rho$ ($\pm 0.01$) & 4.37  & 6.35  & 8.43  & 2.73 \\
    \hline
          & t ($\pm 0.5$) & 34.2  & 5.9   & 34.2  & 2.4 \\

    40 nm & $\sigma$ ($\pm 0.1$) & 1.9   & 2.2   & 1.8   & 1.5 \\

          & $\rho$ ($\pm 0.01$) & 4.21  & 6.12  & 9.22  & 6.04 \\
          \hline
    \end{tabular}
  \label{tab1}
\end{table}

\begin{figure} [ht]
\centering
\includegraphics[scale=0.5]{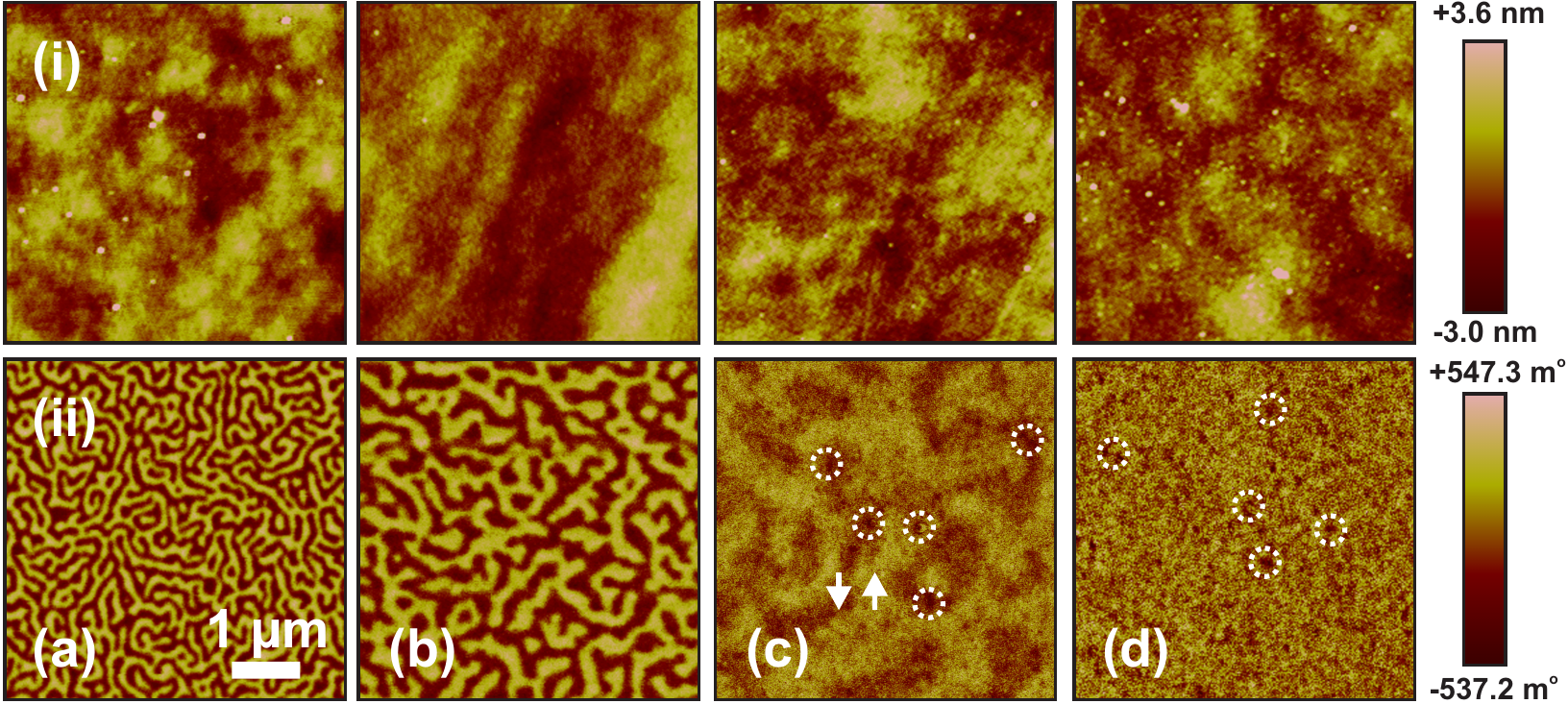}
\caption{(color online) AFM and MFM images of various films. Upper
images (i) represent AFM images and lower (ii) represent
corresponding MFM images. (a) refers for bare TbFe film (UL = 0
nm), (b), (c), and (d) represents for UL = 10, 20, and 40~nm
respectively. The yellow contrast represents up-magnetization
while red contrast as down-magnetization in the MFM images (up-
and down-arrow is represented on the MFM image of UL = 20 nm).
Small white dashed circular dots are some of the point-like
defects highlighted on the surface of the film.} \label{fig:3}
\end{figure}

Ti UL affects the microscopic properties of the sample. Figure
~\ref{fig:3} shows AFM and MFM images of TbFe films grown with
various thicknesses of Ti UL. AFM images revealed that the surface
morphology slightly changed with increase in UL thickness. MFM
images are found to be free from topographic influences. TbFe film
(UL : 0~nm) shows a labyrinth-like stripe pattern as represented
in fig.~\ref{fig:3}-(ii) (a). TbFe system also exhibits a
honeycomb-like structure as reported in literature
~\cite{TbFe_honey}. The yellow, and red colors represent up-, and
down-magnetization respectively. The average domain width and
contrast are enhanced for UL = 10 nm (fig.~\ref{fig:3}$-$(ii)
(b)). These domain-orientations are similar to the case of UL =
0~nm film. In a way, by placing a 10~nm Ti UL enhances the domain
contrast and average domain size in the film. However, domain
contrast becomes weaker for further increment in UL thickness. The
yield in the electron beam evaporation technique is quite high,
so, it is suspected that roughness may increase by increasing the
thickness of the Ti UL. It is observed that the interface
roughness ($\sigma$, Table~\ref{tab1}), and surface roughness
($\approx$ 0.89~nm, Table~\ref{table_example}) of UL = 50~nm are
comparatively higher than the lower UL thicknesses. XRR analysis
(Table~\ref{tab1}) also confirms increase in $\sigma$ value due to
increase in UL thickness. Further increment in UL thickness to
25~nm, stray fields (probing by MFM) that are emanating from the
samples significantly deteriorates resulting in faint maze-like
domains as shown in fig.~\ref{fig:3}-(ii) (c). For UL = 50~nm, a
null out-of-plane contrast, which is observed in MFM image (fig.
~\ref{fig:3}-(ii) (d)).

\begin{figure} [ht]
 \centering
 \includegraphics[scale=0.5]{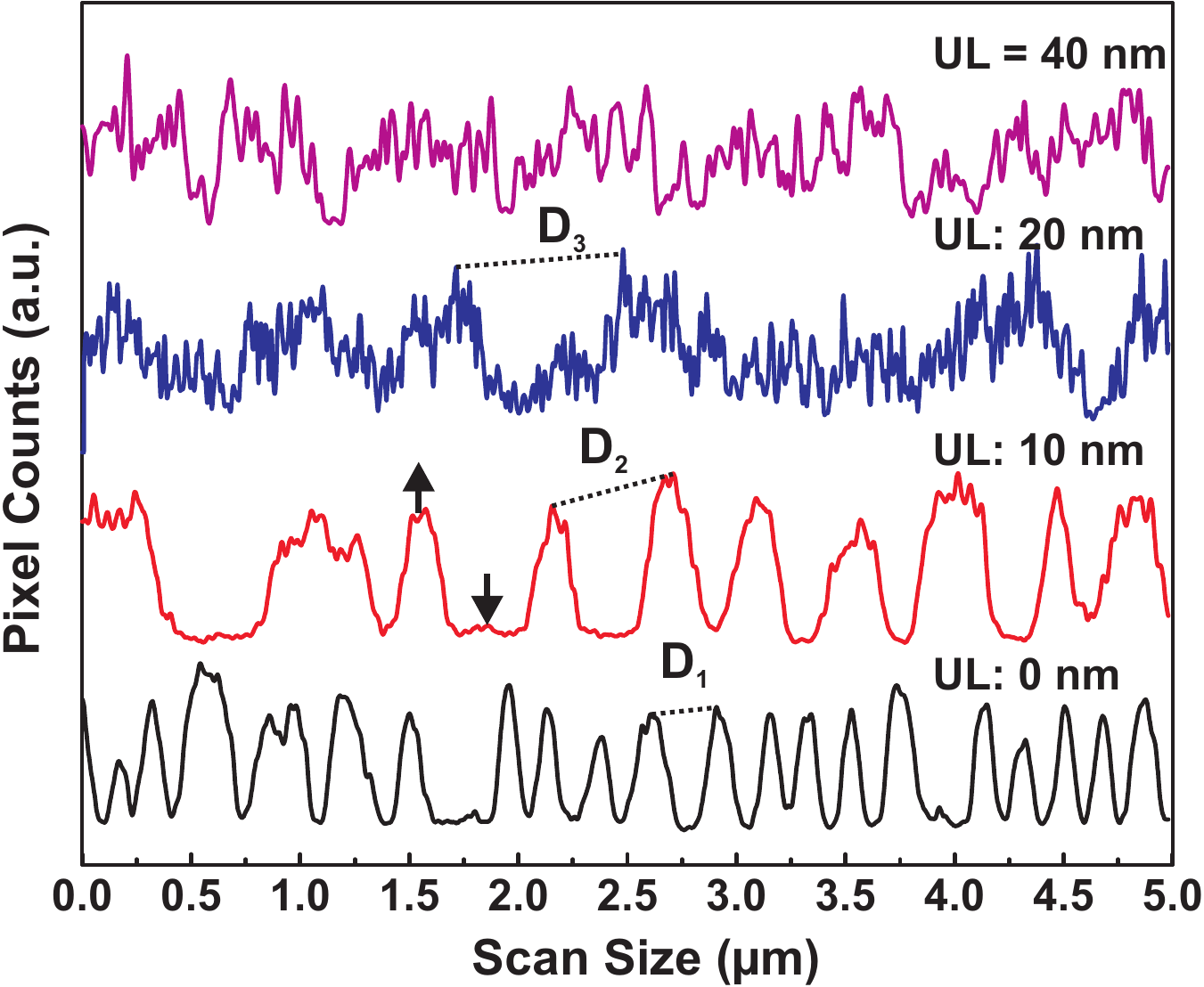}
\caption{(color online) Line scans (at the middle position) on the
MFM images of the various sample leveled as UL = 0, 10, 20, and
40~nm. Up-, and down-magnetization profile are marked in UL =
10~nm line scan. Domain size (D) can be calculated as mentioned in
the line profile, e.g, $D_1$: 0.31 $\mu$m, $D_2$: 0.55 $\mu$m,
$D_3$: 0.78 $\mu$m.} \label{fig:4}
\end{figure}

It was observed that microscopic properties are enhanced for UL
thickness up to 10~nm. Further increment in UL thickness leads to
deterioration of magnetic contrast probably due to the increase of
interface-induced-roughness. The root-mean-squared surface
roughness ($R_q$), areal percentage (\%) of up- and down-domains
are summarized in Table {~\ref{table_example}}. By increasing the
thickness of Ti UL, $R_q$ is found to be increasing. Although
$R_q$ can not significantly describe the details of interfacial
roughness ($\sigma$), still it hints overall roughness of the
sample. For UL = 0~nm, the area of down-domains (60 \%) are more
than up-domains (40 \%). By increasing the UL thickness, \% of
down-domains are decreased compared to \% of up-domains.
Meanwhile, this is observed as UL thickness helped to increase the
average-domain-size, and after a certain thickness, stray fields
that are emanating from these domains become deteriorated. The
domain sizes are extracted by taking a line scan over the MFM
images (fig.~\ref{fig:4}). The sign of up- or down-domains are
marked as up- or down-arrow in the line scan of UL = 10~nm. Domain
size can be calculated by considering two neighboring domains,
mentioned as $D_i$, black dotted lines over the line scans.
Keeping in mind, up to UL = 25~nm, we observed the appearance of
domains. In the case of UL = 50~nm, we did not observe any
domains.

As we observe various thicknesses of Ti UL fascinates the
microscopic domain evolution of the system. So, the reversal
mechanism of these magnetic films would be of fundamental
interest. The magnetization reversals along with Kerr hysteresis
are captured by using the PMOKE technique. MOKE can measure the
change in rotation of polarization of incident light after
reflected from the magnetized surface. Here, the Kerr rotation
(KR) can be detected by even a small change of $\pm$ 5 $m^\circ$.
The MOKE effect is a relativistic quantum effect, which arises due
to the combined effect of two central phenomena. One is
corresponding to the inherent magnetization of the host
sample$-$and$-$other one is the spin-orbit interaction and the
exchange interaction among host atoms. Generally speaking here, KR
values can be considered as additional evidence of correlation of
perpendicular moments in the system.

\begin{figure} [htp]
 \centering
    \includegraphics[scale=0.4]{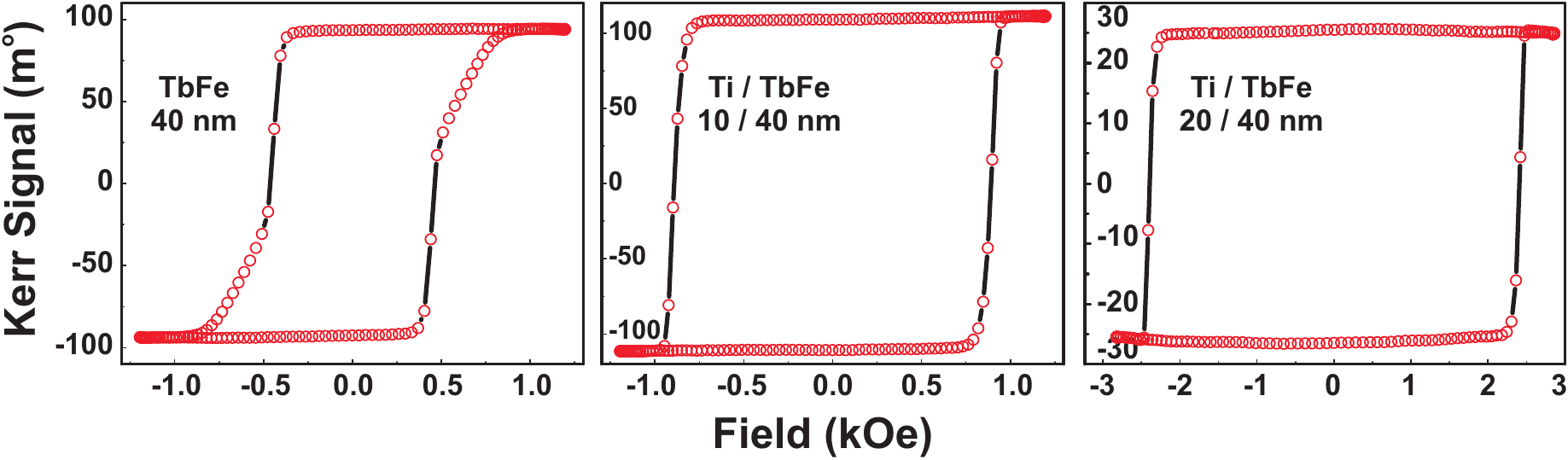}
        \caption{PMOKE measurements of various films.
        \textbf{(i)} Kerr hysteresis of bare TbFe of 100 nm (UL = 0 nm),
        \textbf{(ii)} Kerr hysteresis of UL = 10 nm,
        \textbf{(iii)} Kerr hysteresis of  UL = 20 nm.}
    \label{fig:5}
\end{figure}

Figure~\ref{fig:5}-(i) represents MOKE measurements of 40~nm thick
TbFe film. The Kerr hysteresis (Kerr signal vs Field) shows a
nearly square hysteresis behavior. The $H_c$ and KR were recorded
as 465~Oe and 185 $m^\circ$ respectively. This film is saturated
at a field of $\approx$ 985~Oe. The magnetization reversal is
observed to be initially dominated by nucleation then driven by
domain wall motion. Similarly, we capture both Kerr loops and
reversal domains for both UL = 10, 20~nm (fig.~\ref{fig:5}-(ii),
(iii)). The $H_c$, KR, and saturation field ($H_s$) values are
summarized in the Table~\ref{table_example}. Magnetization
reversals of UL = 10, and 20~nm, are  of similar nature to the
case of bare TbFe 40~nm film. The reversal mechanism initially
started from nucleation-dominated to wall-motion dominated one in
all the cases.

From Table~\ref{table_example}, it is found that the increment of
UL thickness results in gradual increase in OOP $H_c$. This
signifies magnetic moments are still lying perpendicular to the
plane of the sample with increment in UL thickness. OOP $H_c$ of
UL = 20~nm is $\approx$ 60 \% more than in the case of UL = 10~nm.
Whereas, strength of OOP stray field of UL = 20 nm is weakened
compared to UL = 10~nm (fig.~\ref{fig:3}, and corresponding line
scans in fig.~\ref{fig:4}). So, placing a UL ultimately enhance
$H_c$ and $H_s$ values. A more understanding of the samples can be
derived by considering KR values. We observe KR of UL = 10~nm (222
$m^\circ$) shows an improvement over UL = 0~nm (185 $m^\circ$)
(refer Table~\ref{table_example}). This shows a possible
enhancement of correlation, and/or interaction of local magnetic
moments, which is due to 10~nm of Ti UL. While in the case of UL =
20~nm, KR rapidly decreases to 51 $m^\circ$. This may be due to
magnetic moments are intermixed, and/or trapped within the
proximity of interface roughness, which leads to a reduction of KR
values as Kerr signals are only sensitive to magnetic moments. So,
one can correlate these $H_c$ and KR values to the strength of
perpendicular anisotropy.

\begin{table}[h]
\caption{Effect of various thickness of UL (0, 10, 20, and 40~nm)
on the $R_q$, \% of up-/down-domain area ($A_d$), $H_c$, $H_s$,
and Kerr rotation (KR) are summarized here, which have been
extracted from fig.~\ref{fig:3} and fig.~\ref{fig:5}.}
\label{table_example} \centering

\begin{tabular}{|c|c|c|c|c|c|}

\hline

  $t_{UL}$ & $R_q$ & $A_d$ & $H_c$ & $H_s$ & KR\\
 (nm) & (nm) & ($\pm$ 5 \%) & ($\pm$ 5 Oe) & ($\pm$ 5 Oe) & ($m^\circ$)\\
 \hline
 0 & 0.50 & 40/60 & 465 & 985 & 185\\
 \hline
 10 & 0.60 & 48/52 & 890 & 1045 & 222\\
 \hline
 20 & 0.85 & 55/45 & 2410 & 2550 & 51\\
 \hline
 40 & 0.89 & -- & -- & -- & --\\
 \hline
\end{tabular}
\end{table}

The enhancement or deterioration of magnetic and/or microscopic
properties with the increment in thicknesses of Ti UL can be
understood by following conclusive understanding. When a film gets
deposited on the substrate, the system experiences lots of thermal
stress, due to difference in thermal expansion coefficients
between substrate and film, or between films, or mismatch lattice
parameters between adjacent layers. In our case, also, bare
amorphous Tb-Fe film of 40~nm experiences thermal stress, but the
effect would be minimal due to given thickness. However, the
origin of PMA is expected to be more prominent by $Tb$-$Fe$
correlations~\cite{TbFe_origin} which leads to structural
anisotropy in these kinds of amorphous system. In such a case, one
can say that these films are more dominated by magneto-crystalline
anisotropy, as PMA increases by increasing the thickness of the
film~\cite{GdFe_thickness}. When we place Ti UL, Tb-Fe film grows
on UL rather on the Si substrate, microscopic property changes. As
Ti shows crystalline behavior (fig.~\ref{fig:1}), placing as UL
will create a lattice mismatch between Ti and Tb-Fe layers, at the
same time thermal stress on the TbFe layer reduces. However, these
effects are not prominent in our case as the TbFe is inherently
amorphous. In addition to this, there could be another phenomenon,
$d-d$ hybridization, coexist between interface. It is not that
3d-magnetism leads to large exchange interactions (here, $Fe$
atoms) whereas 4f-magnetism drives large magneto-crystalline
anisotropy (here, Tb atoms)~\cite{3d_4f}. The extended length of
the hybridization is more prominent in the ultrathin system, where
interfacial roughness is very nominal. It is also observed that Ti
($d^2$) alloys show a faint-magnetic behavior when it is mixed
with Fe, Co, or Ni (any strong magnetic material)~\cite{Ti_book}.
Up to UL = 10~nm, we observe a nominal roughness $\approx$ 0.60~nm
(Table~\ref{table_example}). The reason for increment in domain
size could be the lateral extent of hybridization of Fe ($d^6$)
and Ti ($d^2$), $d-d$ hybridization, become more prominent than
the extent of interfacial roughness. As the thickness of UL
increases, the extent of interfacial roughness gets dominated over
hybridization effects, as a result, local anisotropy becomes weak
and local moments may get trap in point-like defects present in
the sample. These moments seem to be trapped by the proximity of
interfacial roughness. One such observation was also found in an
ultrathin TbCo system, $<$ 3~nm~\cite{Ti_alloy_PMA}. The reason
for the enhancement of PMA was elucidated due to the appearance of
a rugged interface.

The above understanding of magnetic and microscopic properties of
TbFe films can be (a) by increasing UL thicknesses perpendicular
anisotropy gradually increases regardless of interfacial
roughness, (b) increment in UL thickness beyond a certain
thickness leads to intermixing of magnetic moments with the
proximity of interface roughness, while magnetic moments still
lies perpendicular to the film plane. However strength of OOP
stray fields are decreased. For instance, in UL = 20~nm case, one
can observe with a considerable PMA (as $H_c$ is more) while OOP
stray field is weakened. This contrast mixing properties due to
the UL which is addressed in this study. Magnetometry measurements
$H_c$ of UL = 40~nm has a similar range of value compared to that
of UL = 20~nm.

\section{Summary}

In summary, we have investigated structural, microscopic, and
magnetic properties of magnetic anisotropy in TbFe thin films by
placing a wide range of Ti UL thickness. By increasing the
thickness of UL which results in an increment in interfacial
roughness with the formation of Fe-Ti phases. Here, we observed
that (i) for UL = 10~nm, the possibility of extended $d$-$d$
hybridization dominates over the influence of interfacial
roughness, as a result, the strength of stray fields enhanced,
(ii) for UL = 20~nm and 40~nm, the extent of interfacial roughness
dominates over the hybridization effects as a result stray fields
deteriorated. However, $H_c$ and $H_s$ are gradually increased
with the increment of UL thickness. By placing UL of 20~nm, $H_c$
increases nearly 6 times more than the bare TbFe film, whereas
Kerr rotation decreased by more than 4 times compared to UL of
10~nm. The magnetization reversal studies reveal domain nucleation
followed by domain-wall motion in all the films.

%



\end{document}